\documentclass{article}
\usepackage{graphicx}
\usepackage{geometry}
\begin{document}

\title{Emergence of Self-Sustained Oscillations for SIRS Model on Random Networks} 
\author{M. Ali Saif\\
Department of Physics,\\
Faculty of Education,
University of Amran,
Amran,Yemen\\
masali73@gmail.com}

\maketitle
\begin{abstract}
We study the phase transition from the persistence phase to the extinction phase for the SIRS (susceptible/ infected/ refractory/ susceptible) model of diseases spreading on random networks. By studying temporal evolution and synchronization parameter of this model on random networks, we find that, this model on random networks, shows a synchronization phase in a narrow range of very small values of clustering coefficient. This finding corroborates the conclusion reached in Ref. \cite{kup} that, the clustering coefficient is responsible for the emergence of the synchronization phase in the small world networks.


\end{abstract}
      

\section{Introduction}
Periodicity on occurring epidemics presents a long standing challenge for the formulation of epidemic models. Whenever the reason of occurring that oscillation related to external driving causes, such as
the seasons or arise from the intrinsic dynamics is an open problem. In attempting to simulate that systems, synchronization phase have been observed in some cases such as: seasonal driving \cite{abr}, stochastic dynamics \cite{gus,apa} and a complex network of contacts \cite{kup}. In case of simulating the spread of diseases using the SIRS model, self sustained oscillation has been observed in some cases at specific values of the parameters of those systems such as: SIRS model with distributed delays \cite{gon}, SIRS on small world networks \cite{kup} and SIRS on Scale free networks \cite{pas}. Explanation the reason of occurring that oscillation on those models still also another challenge, needs to reveal.

Kuperman and Abramson \cite{kup} have shown that, SIRS model in the regular one dimensional lattice evolves to a fluctuating endemic state of low infection. However, on small world network and at a finite value of the disorder of the network, they get a transition to self-sustained oscillations in the size of the infected subpopulation. They also found that, that transition specifically occurs where the average clusterization shifts from high to low. Which indicates that, the cause of the phase synchronization is the clustering coefficient. 
Kuperman and Abramson argued that, the clustering coefficient to be the responsible of evident that behavior \cite{kup}. In Ref. \cite{sai} we have proved that, for the SIRS model on the networks and when the recovery time $\tau_R$ is larger than the infection time $\tau_I$, the infection will flow directionally from the ancestors to descendants however, the descendants will be unable to reinfect their ancestors during the time of their illness. That behavior leads him to deduce that, in order to the disease spreads frequently throughout the nodes of the networks, the loops on the network are necessary, which means the clustering coefficient will play an important role in this model. where, clusters tend to spread infection among close-knit neighborhoods. Hence, when the loops are high inside the network, the reinfection occurs in the network at many places and at different times, which looks like a kind of randomness. Whereas when the number of loops are low, reinfection occurs at specific places and times on the network, which looks like as a kind of regularity in occurring the second period of infection.

Inspired of those works, we are going in this work to study the SIRS model on random networks. Whereas, we can control the clustering coefficient on the random networks from very small values, up to 1, we expect to observe the self sustained oscillation in the size of the infected subpopulation of this model when the clustering coefficient is low, similarly as what happen in the case of small world network \cite{kup}.

Erd$\ddot{o}$s-R$\acute{e}$nyi (ER) \cite{alb} random network is defined in the following way: one start with $N$ isolated nodes, every pair of nodes being connected with probability $p$, leading to a
Poisson degree distribution $P(k)=\exp^{-\left\langle k\right\rangle}\left\langle k\right\rangle^k/k\!$ with average degree 
\begin{eqnarray}
\left\langle k\right\rangle=(N-1)p
\end{eqnarray}

The total number of edges $L$ is a random variable with the expectation value $\left\langle L\right\rangle=p\frac{N(N-1)}{2}$. The clustering coefficient of a random graph is
\begin{eqnarray}
 C_{rand}=p=\frac{\left\langle k\right\rangle}{N}
\end{eqnarray}

\section{Epidemic model}
SIRS epidemic model on the networks is defined as follows \cite{kup,gade}:  
if we consider a lattice of $N$ nodes. Each node connected to its $k$ neighbors. The nodes can exist in one stage of three states, susceptible $(S)$, infected $(I)$ and refractory $(R)$. 
Susceptible node can pass to
the infected state through contagion by an infected one.
Infected node pass to the refractory state after an
infection time $\tau_I$. Refractory nodes return to the
susceptible state after a recovery time $\tau_R$. The contagion
is possible only during the $S$ phase, and only by a
$I$ node. During the $R$ phase, the nodes are immune
and do not infect. The system evolves with discrete time
steps. Each node in the network is characterized by a
time counter $\tau_i(t)= 0, 1, ..., \tau_I + \tau_R\equiv \tau_0$ , describing its
phase in the cycle of the disease. The epidemiological
state $\pi_i$ of the node $(S, I, or R)$ depends on the phase
in the following way:
\begin{eqnarray}
\pi_i(t)&=S & \mbox{if} \tau_i =   0,\nonumber\\
\pi_i(t)&=I & \mbox{if} \tau_i \in (1,\tau_I),\nonumber\\
\pi_i(t)&=R & \mbox{if} \tau_i \in (\tau_I +1,\tau_0)
\end{eqnarray}
The state of a node in the next step depends on its current phase in the cycle. A susceptible node stays as such, at $\tau= 0$, until it becomes infected. Once infected, it goes
(deterministically) over a cycle that lasts $\tau_0$ time steps. During the first $\tau_I$ time steps, it is infected and can potentially transmit the disease to a susceptible neighbor.
During the last $\tau_R$ time steps of the cycle, it remains in state $R$, immune and not contagious. After the cycle is complete, it returns to the susceptible state.

\section{Numerical Results}

In this work, we consider a synchronous update of a system of $N$ nodes on a random network, in which any link connects two nodes exists with probability $p$, hence each node is connected on average to $\left\langle k \right\rangle$ neighbors. We assume that, at each time step an infected node infects each of its neighbors with probability $\lambda$. Infected nodes remain as such for $\tau_I$ time steps, after which they become immune for $\tau_R$ time steps. The probability $T$ for the infected node $i$ to infect one of its neighbors during the time of its illness is: \cite{new,pars,cohen}:
\begin{eqnarray}
T=\left[1-(1-\lambda)^{\tau_I}\right]
\end{eqnarray}

We have performed extensive numerical simulations of the described model on ER random network. For system of $N=10^4$ with an initial fraction of $n_{inf}(0)=0.1$ infected nodes, and the rest are susceptible, Fig. 1 shows part of three time series displaying the fraction of infected nodes $n_{inf}(t)$ in the system. The $200$ time steps shown are representative of the stationary state of system. Fig. 1(right) shows three curves correspond to systems with different values of connection probability $p$: $p=0.0004$ (bottom), $p=0.0006$ (middle) and $p=0.0008$ (top). Whereas Fig. 1(left) shows three curves correspond to systems with different values of connection probability $p$: $p=0.001$ (bottom), $p=0.0014$ (middle) and $p=0.0016$ (top). All the curves in Fig. 1 for the case when the infection time is $\tau_I=7$, recovery time is $\tau_R=10$ and infection probability $\lambda=0.1$. Fig.1 shows clearly a transition from fluctuating endemic state to an oscillatory one followed by fluctuating endemic situation. At $p=0.0004$ (this value of $p$ corresponds to average degree $\left\langle k\right\rangle=4$, Eq. 1) and $p=0.0006$ (average degree $\left\langle k\right\rangle=6$) the stationary state is a fixed point, with fluctuations. Self sustained oscillation appears clearly at the values of $p=0.0008$ (average degree $\left\langle k\right\rangle=8$) and $p=0.001$ (average degree $\left\langle k\right\rangle=10$). As the value of $p$ increases the amplitude of oscillation starts decline as that clear at the value of $p=0.0014$ ($\left\langle k\right\rangle=14$). System again reaches the fixed point with fluctuation when $p=0.0016$ ($\left\langle k\right\rangle=16$).

\begin{figure}[htb]
 \includegraphics[width=70mm,height=60mm]{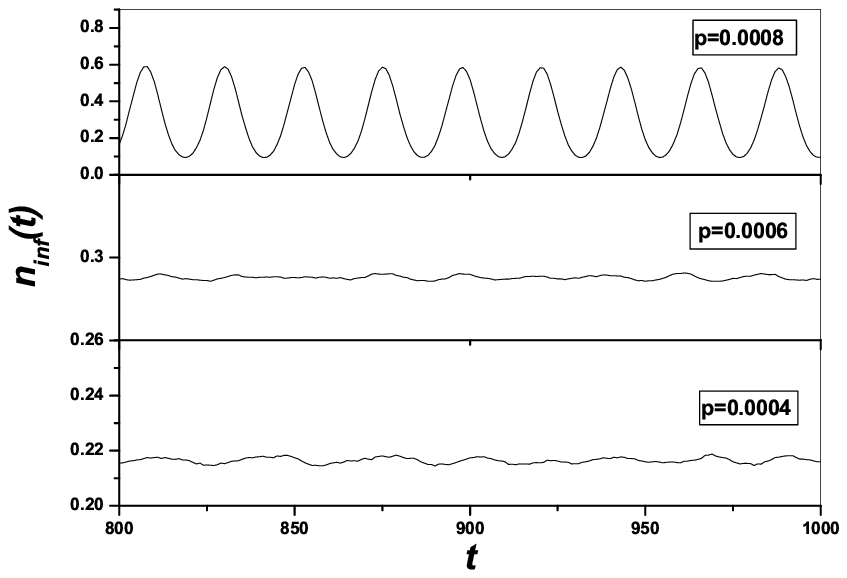}
 \includegraphics[width=70mm,height=60mm]{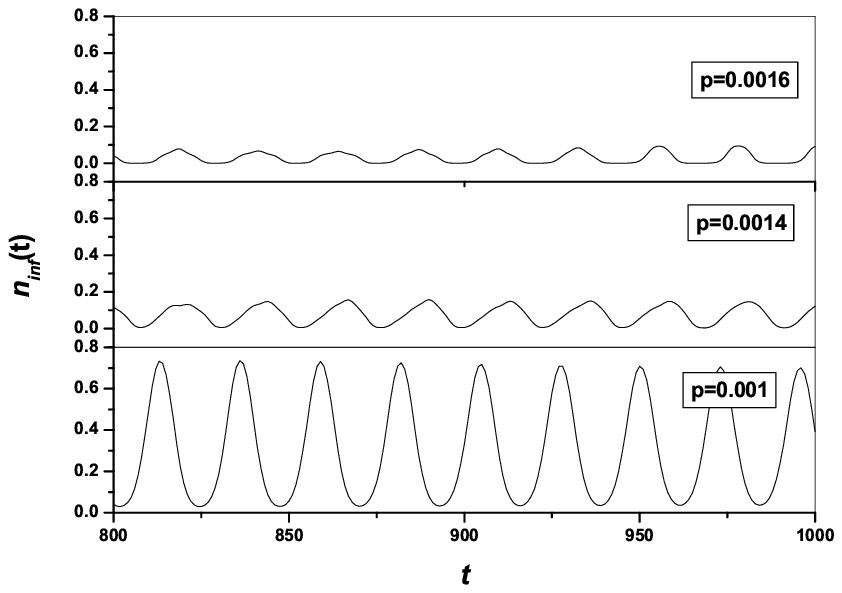}
\caption{ Density of infectious nodes as function of time for different values of $p$, as shown in the legend. Other parameters are $N=10^4$, $\tau_I=7$, $\tau_R=10$, $\lambda=0.1$ and $n_{inf}(0)=0.1$.}
 \end{figure} 

It is evident from Fig. 1 that, in narrow range of the values of $p$ specifically from $p=0.0008$ ($\left\langle k \right\rangle=8$), up to $p=0.0014$ ($\left\langle k \right\rangle=14$), ER random network has a large amplitude self-sustained oscillation. Oscillation almost periodic with a very well defined period and small fluctuations in amplitude similarly to the situation on small world network \cite{kup}. Periodic time of that oscillation approximately equal to $23$ time steps, which is slightly longer than $\tau_0=\tau_I+\tau_R$, since it includes the average time that a susceptible node remains at state S, before being infected.

\begin{figure}[htb]
 \includegraphics[width=70mm,height=60mm]{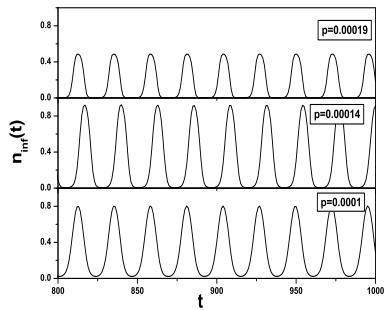}
\caption{ Density of infectious nodes as function of time for different values of $p$, as shown in the legend. Other parameters are $N=10^5$, $\tau_I=7$, $\tau_R=10$, $\lambda=0.1$ and $n_{inf}(0)=0.1$.}
 \end{figure} 

For the case when the system size is $N=10^5$, we show in Fig. 2, and for the same values of the parameters in Fig. 1, three curves correspond to system with different values of connection probability $p$: $p=0.0001$ (bottom), $p=0.00014$ (middle) and $p=0.00019$ (top). We averaged over $5$ configurations. It is clear here that, oscillation becomes better and reaches its highest value when $p=0.00014$ ($\left\langle k \right\rangle=14$). Oscillation we observed here also is in narrow range of vales of $p$, precisely from $p=0.00008$ ($\left\langle k \right\rangle=8$), up to $p=0.00019$ ($\left\langle k \right\rangle=19$).   
  
The periodic oscillation in the density of infected nodes on the network indicates to a spontaneous synchronization of a significant fraction of the nodes on that network. In order to quantify that synchronization, we use the synchronization parameter \cite{kur}, which is defined as
\begin{eqnarray}
\sigma(t)=\left|\frac{1}{N}\sum_{j=1} ^{N} \exp^{i\phi_j(t)}\right|
\end{eqnarray}

Where $\phi_j=2\pi(\tau_j-1)/\tau_0$ is a geometrical phase corresponding to $\tau_j$. We does not include the sum in the previous equation, the case when $\tau=0$ \cite{kup}. When the nodes on the network are not synchronized the phases are spread widely in the cycle and the complex numbers $\exp^{i\phi}$ are correspondingly spread in the unit circle. In this case $\sigma$ is small. On the other hand, when a significant part of the nodes are synchronized in the cycle, $\sigma$ is large.

Fig.3 shows the average value of synchronization parameter $\left\langle \sigma\right\rangle$ of the system as a function of connection probability $p$, for the case when the system size is $N=10^4$ (left) and $N=10^5$ (right). Each point on the figure corresponds to a time average of $1000$ time steps after we discard $2000$ time steps, and a subsequent average over $20$ configurations of the network when $N=10^4$ and $5$ configurations of the network when $N=10^5$. Other parameters are $\tau_I=7$, $\tau_R=10$, $\lambda=0.1$ and $n_{inf}(0)=0.1$.
 
It is clear that, synchronization phase in this case exists at very small values of clustering coefficient of random networks. The point which we assert here is this, synchronization phase in this system appears approximately at $\left\langle k \right\rangle\approx 8$ (which is $p\approx0.0008$ when $N=10^4$ and $p\approx 0.00008$ when $N=10^5$). This value of $\left\langle k \right\rangle$ is larger than the critical point $p_c\approx1/N=\left\langle k\right\rangle=1$, where the giant component emerges on the random network. The value of $\left\langle k \right\rangle$ at which synchronization phase starts appearing almost is the value of $\left\langle k \right\rangle$ at which the isolated node on random network is absent and the random network becomes connected ($\left\langle k\right\rangle > \ln N(p> \frac{\ln N}{N})$). Synchronization on the random network as Fig. 3 shows, disappears almost at $\left\langle k\right\rangle\approx17$ when $N=10^4$ and at $\left\langle k\right\rangle\approx19$ when $N=10^5$.

\begin{figure}[htb]
 \includegraphics[width=70mm,height=60mm]{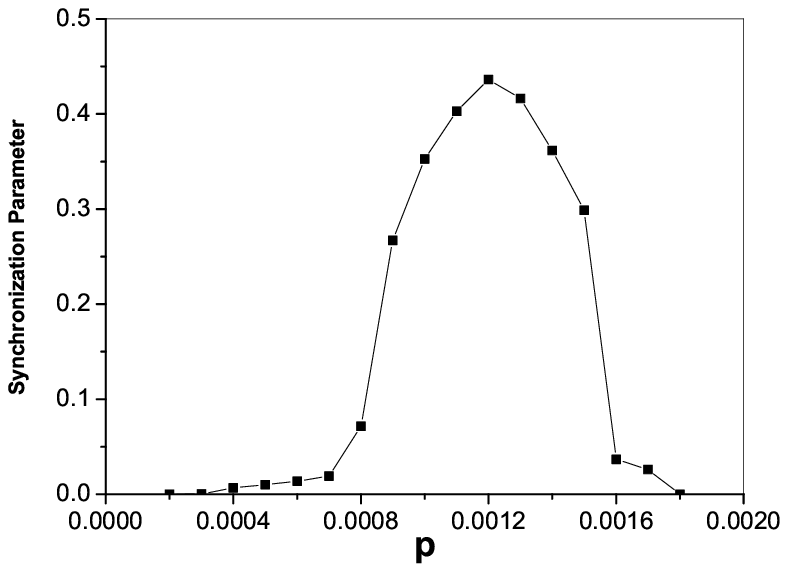}
 \includegraphics[width=70mm,height=60mm]{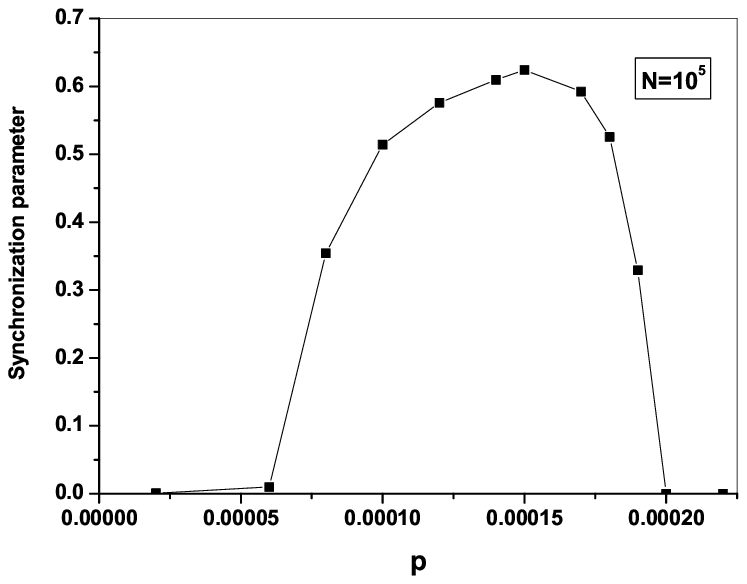}
\caption{Average synchronization parameter $\left\langle \sigma \right\rangle$ of the system as a function of connection probability $p$, when $N=10^4$ (left) and $N=10^5$ (right) for the case of $\lambda=0.1$. Each point corresponds to a time average of $1000$ time steps, and a subsequent average over $20$ configurations of the network when $N=10^4$ and $5$ configurations of the network when $N=10^5$. Other parameters are $\tau_I=7$, $\tau_R=10$, and $n_{inf}(0)=0.1$.}
 \end{figure}

The results obtained here confirm what we had previously speculated about the possibility of synchronization at low values of clustering coefficient for this model on random networks. This result also confirms that, the important role played by the clustering coefficient in the occurrence of oscillation in this model on the networks. The results obtained here are also consistent with the results obtained in Ref. \cite{kup} on the responsibility of that parameter in the appearance of the synchronization state in such systems. In addition to that, we can confirm the importance of this parameter in the occurrence of the synchronization state from Ref. \cite{sai}, which emphasizes the need for the loops in the networks in order to the disease to spread frequently throughout the nodes of the networks. Whereas, clusters tend to spread infection among close-knit neighborhoods \cite{lew}. We speculate that, whenever the value of clustering coefficient is high the network is highly clustered  and the loops exist at many places on the network, hence the next period of infection will happen at many places on the network and at any time, which will look like as a kind of randomness (in space and time) in the occurring the next generation of infection. However, when the clustering coefficient becomes lower, which means the number of triangular loops on the network also will become lower, the reinfection will be localized where those loops exist, consequently the next period of infection (on the average) will happen at specific place and time on the network. Which will look like as a kind of regularity in occurring the second period of infection.

\section{Conclusion} 
We have studied the spreading of infectious diseases for the SIRS model on the random networks. Simulation results of this model on random network show a synchronization phase in narrow range of the values of clustering coefficient, similarly to what has been observed in small world network \cite{kup}. This finding confirms what the authors conclude in Ref. \cite{kup} about responsibility of the clustering coefficient parameter in occurring the synchronization for this model on networks.

 \end{document}